\documentclass{PoS}
\pdfoutput=1
\usepackage{epsfig,graphicx}
\usepackage{amssymb}

\newcommand{\beq}{\begin{equation}}
\newcommand{\eeq}{\end{equation}}
\newcommand{\be}{\begin{equation}}
\newcommand{\ee}{\end{equation}}
\newcommand{\ba}{\begin{array}}
\newcommand{\ea}{\end{array}}
\newcommand{\beqa}{\begin{eqnarray}}
\newcommand{\eeqa}{\end{eqnarray}}
\newcommand{\bea}{\begin{eqnarray}}
\newcommand{\eea}{\end{eqnarray}}
\newcommand{\lsim}{\stackrel{<}{_\sim}}
\newcommand{\gsim}{\stackrel{>}{_\sim}}

\newcommand{\cA}{{\cal A}}
\newcommand{\cO}{{\cal O}}
\newcommand{\cB}{{\cal B}}
\newcommand{\cH}{{\cal H}}

\newcommand{\no}{\nonumber}

\title{The Challenges of Flavour Physics}

\ShortTitle{The Challenges of Flavour Physics}

\author{\speaker{Gino Isidori}\thanks{Work supported by the 
Technische Universit\"at M\"unchen - Institute for Advanced Study, 
funded by the German Excellence Initiative.} \\
        INFN, Laboratori Nazionali di Frascati, Via E. Fermi 40
I-00044 Frascati, Italy.\\
        E-mail: \email{Gino.Isidori@lnf.infn.it}}
  
\abstract{
The open problems and the most recent developments in flavour physics
are briefly reviewed.
Particular attention is devoted to the current ``anomalies'' in the CKM picture and 
their possible interpretation in  beyond-the-Standard-Model frameworks.

\hskip 12 true cm.

To the memory of Nicola Cabibbo.
}

\FullConference{35th International Conference of High Energy Physics - ICHEP2010,\\
		July 22-28, 2010\\
		Paris France}

\begin{document}

\section{Introduction: the big challenges in flavour physics}

A lot of progress has been made since 1963, when the first building 
block of what we now call flavour physics was laid down~\cite{Cabibbo:1963yz}.
The mechanism of flavour mixing has been tested with high accuracy
in the quark sector, where all flavour-violating phenomena
seems to be
well described by the Standard Model (SM)
Yukawa interaction with three generations of quarks~\cite{Kobayashi:1973fv}.
Flavour mixing 
has been observed also in the neutrino sector, indicating the
existence of a non-vanishing neutrino mass matrix which cannot 
be accommodated within the SM. 


Despite this important progress, the origin of flavour
is still a mystery. Our ``ignorance'' can be summarized by the following two open questions:
\begin{itemize}
\item{} What determines the observed pattern of masses and mixing angles of quarks
and leptons?
\item{} Which are the sources of flavour symmetry breaking accessible at low energies?
Is there anything else beside the SM Yukawa couplings and the neutrino mass matrix?
\end{itemize}
The attempts to answer the first question are typically based on 
the introduction of a non-trivial flavour dynamics at some high scale
(see e.g.~\cite{Froggatt:1978nt}).
The new dynamics can be 
associated to Abelian or non-Abelian continuous symmetries (see e.g.~\cite{Lalak:2010bk}
for a recent discussion)
or, as suggested by the neutrino sector, to a discrete symmetry
(see e.g.~\cite{Altarelli:2010gt}).
Alternatively, in models with extra space-time dimensions, the 
flavour hierarchy could be an infrared-property associated to the different
localization of the fermion profiles in extra dimensions (see e.g.~\cite{Grossman:1999ra}).
In all cases it is quite easy to reproduce the observed mass matrices in terms of
a reduced number of free parameters, while it is difficult to avoid problems 
with Flavour Changing Neutral Currents (FCNCs) unless some amount of 
fine-tuning is introduced. 
Most important, it is not easy to make progress in answering 
this question without knowing the ultraviolet completion of the model.

Answering the second question is more easy: it is mainly a question 
of precision, both on the theory and on the experimental side.
Despite the phenomenological success of the SM, we have 
clear indications that this theory needs an ultraviolet completion. 
The most realistic  proposals point toward the existence 
of new degrees of freedom at the TeV scale, possibly accessible at the 
high-$p_T$ experiments at the LHC. In this perspective the second question 
above is related to the flavour structure of these new degrees 
of freedom: is flavour symmetry breaking at the TeV scale fully controlled 
by the SM Yukawa couplings and the neutrino mass matrix? 
As far as the effects in low-energy observables are concerned, this question can 
be formulated within a general effective theory approach and, to a large extent,
it is independent from the ultraviolet dynamics (see e.g.~\cite{Isidori:2009px}).
Following this approach we have already learned a lot about the 
possible flavour structure of physics beyond the SM.
As I  will illustrate in the rest of this talk, 
we could learn even more improving the precision in 
selected low-energy observables.

\subsection{Present status of CKM fits: the global picture}
\label{sect:MFV}

\begin{figure}[t]
\begin{center}
\includegraphics[width=65mm]{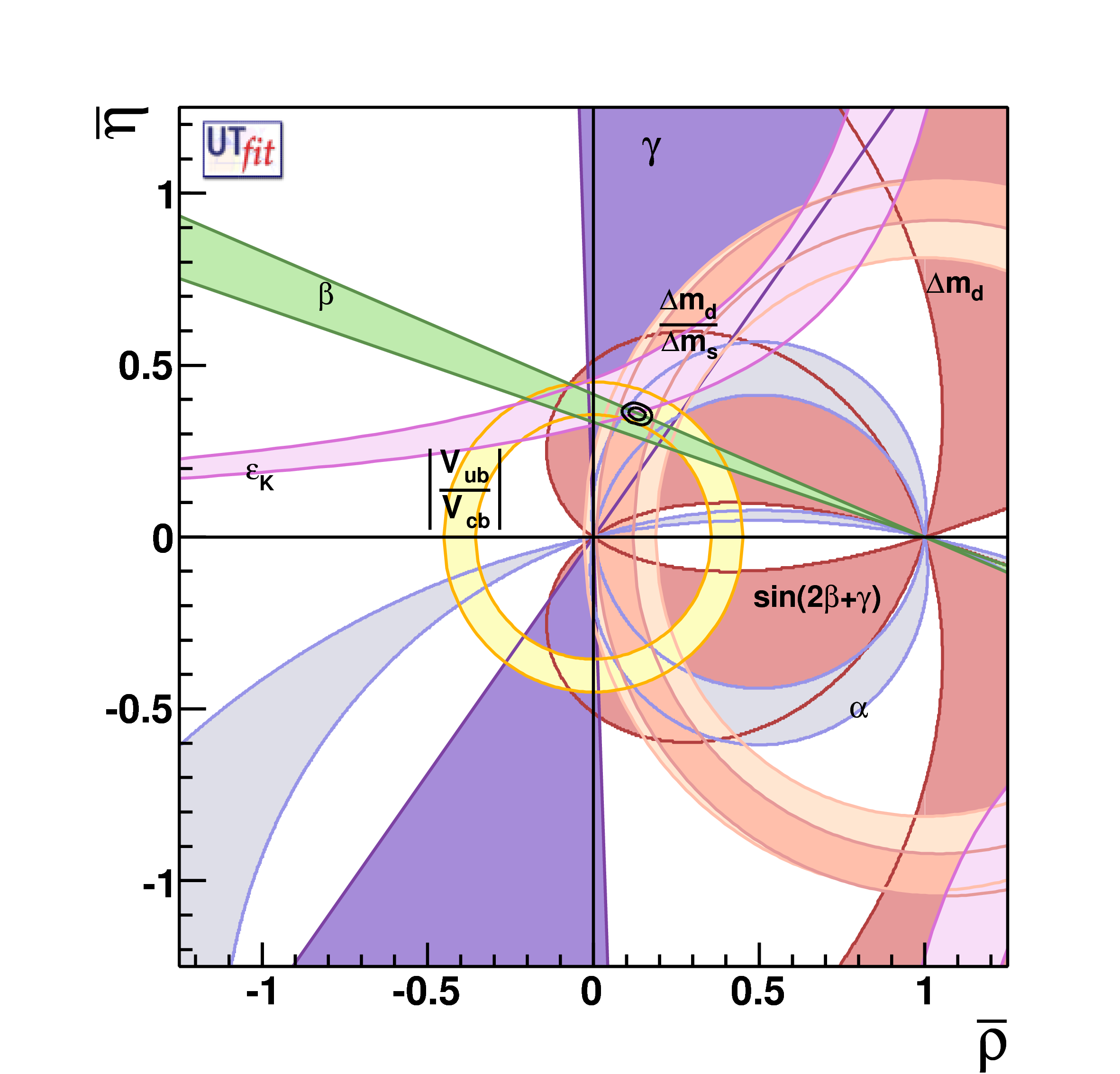}
\includegraphics[width=65mm]{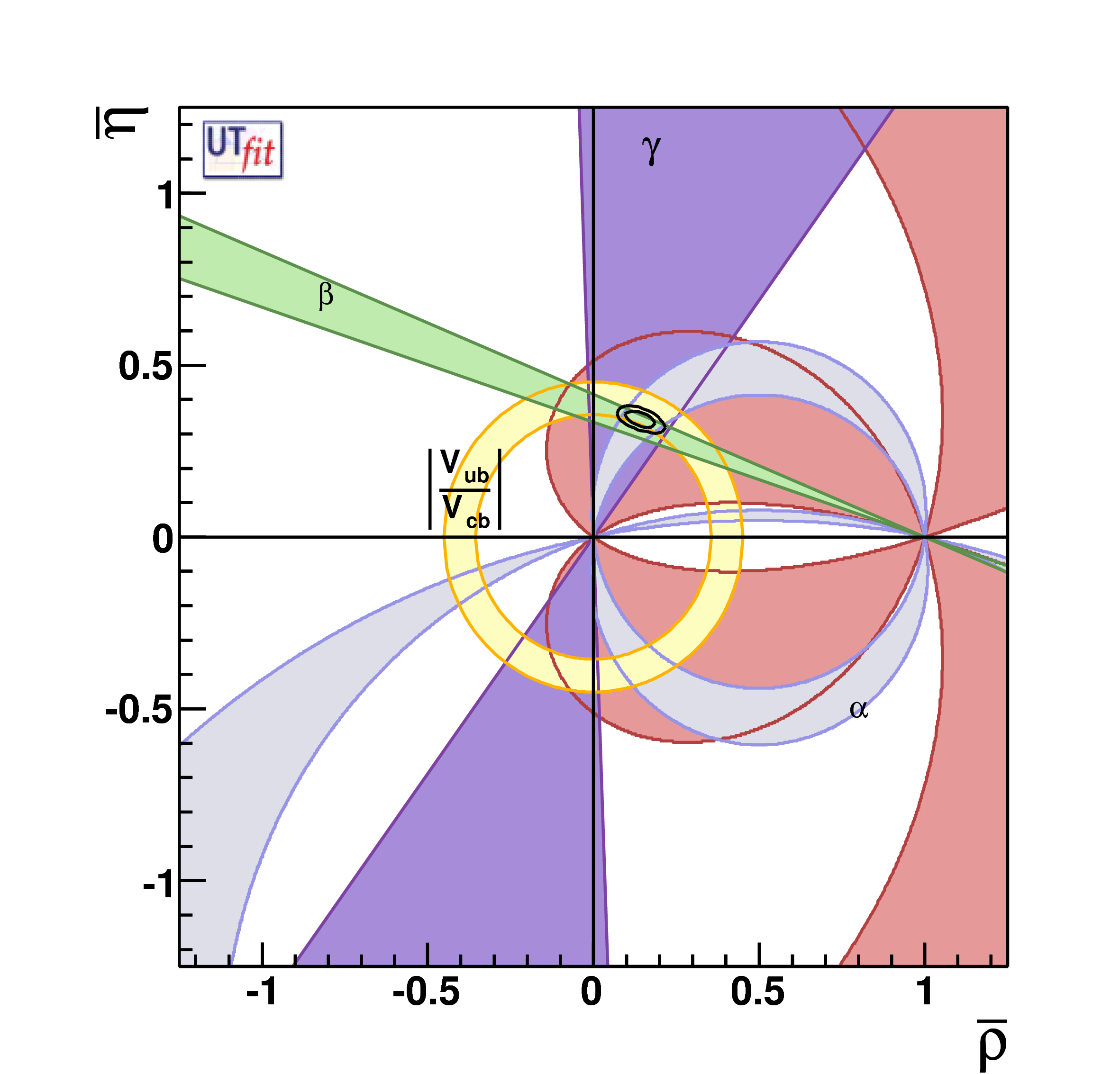}
\caption{\label{fig:UTfits} Fits of the CKM unitarity 
triangle from Ref.~~\cite{Bevan:2010gi}. Left: SM case.
Right: CKM constraints in  extensions of the SM satisfying 
the MFV hypothesis (universal unitarity triangle~\cite{Buras:2000dm}). 
}
\end{center}
\end{figure}

The good overall consistency of the experimental constraints appearing in the so-called
CKM fits (see~\cite{Bevan:2010gi,CKMfitter} for updated results) 
indicate that there is not much room for new sources of flavour
symmetry breaking accessible at low energies (see Fig.~\ref{fig:UTfits}). 
The success of the SM in describing flavour mixing is also confirmed by a series of other 
observations. Two notable examples are: i) the agreement between the SM prediction and 
the experimental determination of $\cB(B\to X_s\gamma)$, where both theory and experimental
errors are below the $10\%$ level~\cite{Misiak:2010dz,Hurth:2010tk}; 
ii)~the test of the CKM unitarity relation 
$|V_{id}|^2+|V_{us}|^2+|V_{Yb}|^2=1$, which is presently probed below the per-mil level~\cite{Antonelli:2010yf}.
All these precise tests can be translated into stringent bounds on physics beyond the SM
(see~\cite{Isidori:2010kg}). These bounds allow us to conclude that 
new flavor-breaking sources comparable and not aligned to the SM Yukawa couplings are 
excluded for new degrees of freedom at the TeV scale. 

The absence of large deviations form the SM in flavour-changing processes, 
together with the need of new degrees of freedom at the TeV scale to stabilize
the SM Higgs sector, is the 
main motivation for the so-called Minimal Flavour Violation (MFV) hypothesis. 
Under this assumption, the SM Yukawa couplings are the only flavour
symmetry breaking terms also beyond the SM~\cite{D'Ambrosio:2002ex}. 
More precisely, in the limit of vanishing quark Yukawa couplings the effective Lagrangian 
describing both SM and new degrees of freedom is invariant under the global 
quark flavour symmetry $SU(3)_{Q_L}\times SU(3)_{D_R} \times SU(3)_{U_R}$~\cite{Chivukula:1987py}.
Employing this hypothesis non-standard contributions in 
flavour-violating transitions turn out to be suppressed to 
a level consistent with experiments even for New Physics (NP) in the 
TeV range~\cite{D'Ambrosio:2002ex}.  The MFV hypothesis provides the technical
tool to address the second question 
listed in the Introduction: if MFV holds, then there are no other sources of
flavour symmetry breaking accessible at low energies. 
Two comments are in order: 
\begin{itemize}
\item{}
The MFV ansatz is quite successful on the phenomenological side; however, 
it is unlikely to be an exact property of the model valid to all scales. 
Despite some recent attempts to provide a dynamical justification 
of this symmetry-breaking ansatz (see e.g.~\cite{Zwicky:2009vt,Albrecht:2010xh}), 
the most natural possibility 
is that MFV is only an accidental low-energy  property of the theory~\cite{Grinstein:2010ve}.
It is then very important to search for possible deviations (even if tiny) 
from the MFV predictions.
\item{}
Even if the MFV ansatz holds, it does not necessarily 
imply small deviations from the SM predictions in all flavour-changing phenomena. 
The MFV ansatz can be implemented in different ways. For instance, in models with 
two Higgs doublets we can change the relative normalization of the two Yukawa 
couplings~\cite{D'Ambrosio:2002ex}, we can decouple the breaking of CP invariance 
from the breaking of the $SU(3)_{Q_L}\times SU(3)_{D_R} \times SU(3)_{U_R}$
quark-flavour group~\cite{Kagan:2009bn} and, in models with strong dynamics at the TeV scale,
we can consider operators with a large number of Yukawa insertions~\cite{Kagan:2009bn}. 
All these variations leads to different and well defined patterns 
of possible deviations from the SM that we have just started to investigate.
\end{itemize}

\section{Recent phenomenological challenges to the CKM picture}

As discussed in the previous Section, the overall picture of quark flavour mixing 
shows a good consistency with the SM predictions. However, looking more closely, 
there are a few cases where the agreement is not so good. The most interesting 
``anomalies'' that have emerged in the last few years are: 
i) the  $\sin 2\beta$ tension in the CKM fit; ii) 
CP violation (CPV) in $B_s$ mixing; iii) the $B\to \tau\nu$ branching ratio.
These (minor) deviations from the CKM picture 
 are particularly interesting since on the one hand the 
NP contributions are small and compatible with the absence of large NP signals 
in other observables, on the other hand the theory errors of the SM predictions 
are small and/or can be systematically improved in the near future.

\subsection{The $\sin 2\beta$ tension in the CKM fit}

\begin{figure}[t]
\begin{center}
\includegraphics[width=65mm]{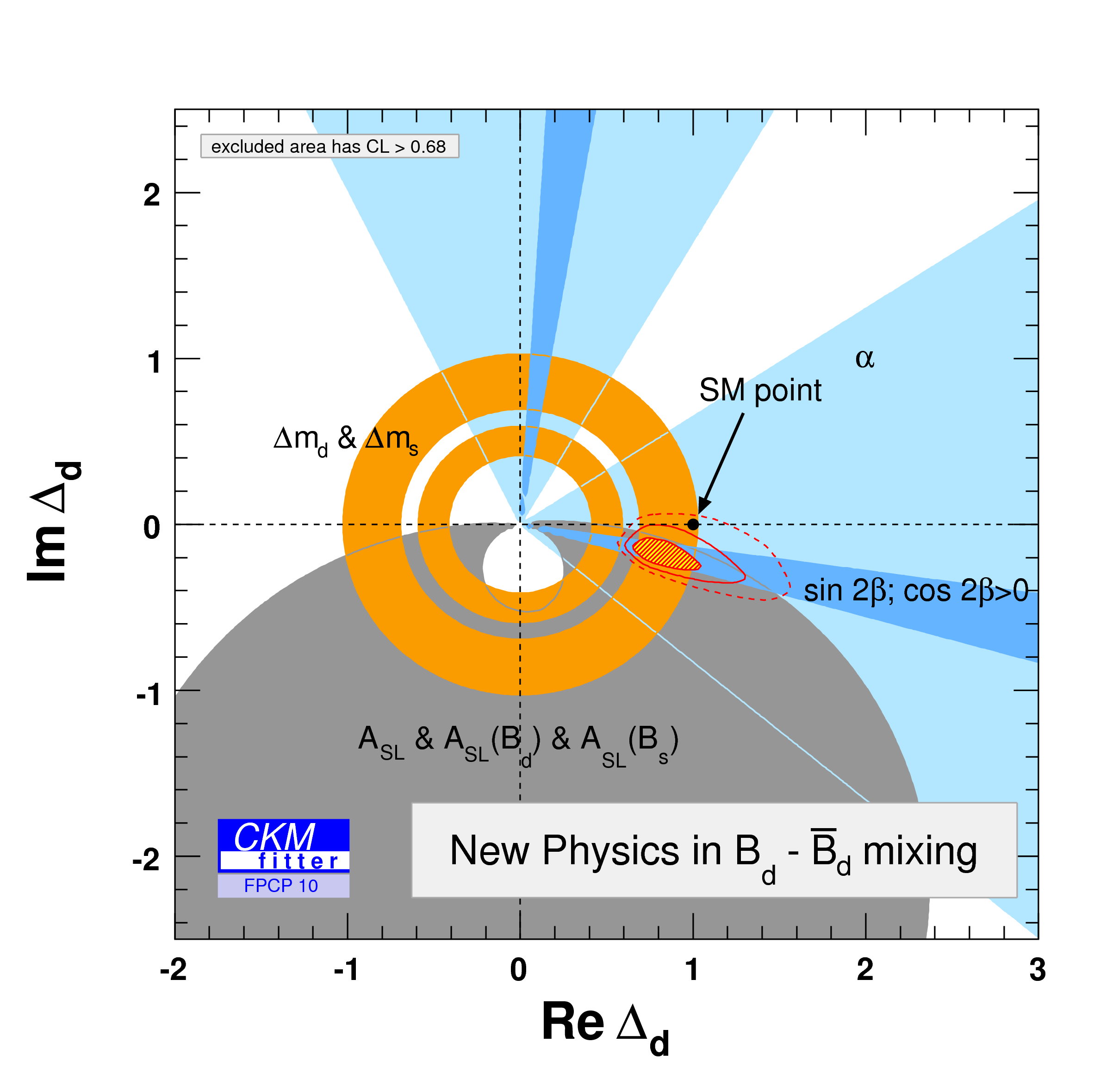}
\includegraphics[width=65mm]{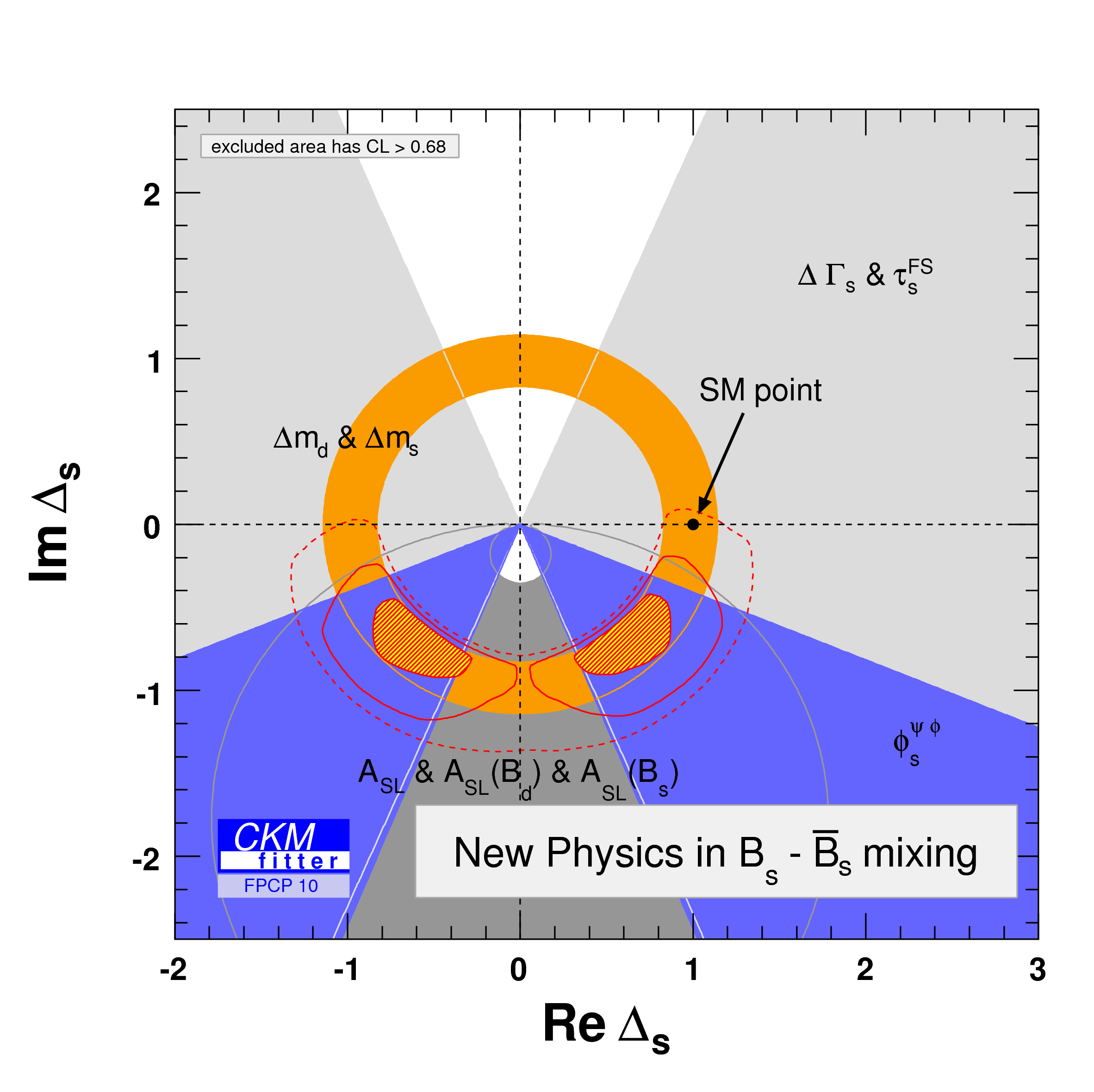}
\caption{\label{fig:DeltaDs}
Constraints on NP in the (Re$\Delta_q$,~Im$\Delta_q$) planes~\cite{Lenz:2010gu}. In both plots 
the three red curves (filled, plain, dashed) correspond to 1, 2 and $3\sigma$ 
contours. The right plot (NP in $B_s$ mixing) includes the measurement of 
$a^b_{\rm SL}$ by D0 but not the very recent D0(6.1 $fb^{-1}$) and CDF(5.2 $fb^{-1}$) 
data on $S_{\psi\phi}$.}
\end{center}
\end{figure}

Within the SM the time-dependent CP asymmetry in $B_d \to \psi K$ (denoted $S_{\psi K}$)
is expected to be equal to $\sin(2\beta)$ (up to corrections below the $1\%$ level),
where $\beta=\arg[-V_{td}^*V_{tb}/(V_{cd}^*V_{cb})]$ (green band in Fig.~\ref{fig:UTfits} left).
At present the experimental determination of 
$S_{\psi K}$ is about $2\sigma$ lower that the 
indirect determination of $\sin(2\beta)$ from other 
observables~\cite{Bevan:2010gi,CKMfitter}.
This problem, which has been noted first in~\cite{Lunghi:2008aa,Buras:2008nn}, 
could be explained by a NP contribution to the $B_d$ mixing amplitude.
A detailed statistical analysis of this possibility has recently performed 
in~\cite{Bevan:2010gi,Lenz:2010gu}. Following the notation of~\cite{Lenz:2010gu},
the NP contribution to $B_q$ mixing amplitudes are parametrized by the complex 
quantities $\Delta_q$, defined by  
\be
 M_{12}^q  = \Delta_q \times  (M_{12}^q)_{\rm SM}~,
\qquad\qquad  2 M_{B_q} M_{12}^q =\langle \bar B_d |\cH_{\rm eff}| B_d \rangle^*~,
\qquad\qquad q=d,s~,
\ee
such that the SM limit is recovered for $\Delta_q=1$. Flavour observables are then 
fitted assuming NP contributes significantly only to  $B_d$ and $B_s$ mixing. 
The best fit value of $\Delta_d$ is shown in the left plot in Fig.~\ref{fig:DeltaDs}.
As can be seen, the SM point is more than $2\sigma$ off the best fit
value (a similar conclusion has been obtained also in~\cite{Bevan:2010gi}). 

While this deviation from the SM is not statistically significant yet, 
the magnitude of the effect is tantalizing for many realistic NP models.
In order to improve the significance of this test we would need to decrease 
the error on the indirect determination 
of $\sin(2\beta)$ via better determinations of $|V_{ub}|$, $\gamma$, and the 
``$\epsilon_K$ band''.  As far as $\epsilon_K$ is concerned, there is a potential warning 
because of an irreducible theoretical error associated to long-distance effects. 
However, as recently show in~\cite{Buras:2010pza}, this error do not exceed the $2\%$ level and 
it is still largely sub-dominant.  

Note that the best fit value of $\Delta_d$ indicates a non-vanishing NP phase in $B_d$ 
mixing. If statistically significant, this would unambiguously indicates
the presence of new CPV phases in addition to the
unique non-trivial CPV phase in the CKM matrix.

\subsection{CPV in $B_s$ mixing}   

Within the SM the $B_s$ mixing amplitude ($M_{12}^s$) has a vanishing CPV phase 
relative to all the leading (tree-level)  $B_s$  decay amplitudes. The relative phase 
between $M_{12}^s$ and the nonleptonic decay amplitude
$b\to c\bar c s$ can be determined experimentally via the $B_d \to \psi K$ 
time-dependent CP asymmetry ($S_{\psi\phi}$). Moreover, the relative 
phase between $M_{12}^s$ and $\Gamma_{12}^s$ can be  determined 
by the semileptonic asymmetry $a^s_{SL}$
(see~\cite{Lenz:2010gu,Ligeti:2010ia,Borissov} for more details). 
The null result expected in both cases is a very clean test of the SM.
  
Since 2008 both CDF and D0 have started to provide a series of 
measurements $S_{\psi\phi}$, while only very recently D0 has provided the first determination 
of $a^s_{SL}$. More precisely, D0 has measured a linear combination of $a^s_{SL}$
and $a^d_{SL}$ ($a^b_{SL} \approx 0.5 a^s_{SL}+0.5 a^d_{SL}$) that, 
combined with $B$-factories data on  $a^d_{SL}$, allow us to determine 
$a^s_{SL}$~\cite{Abazov:2010hv}. The central values of the various
measurements tend to favor a non-vanishing $B_s$ mixing phase, such 
that $S_{\psi\phi}$ is positive and of order 1.
The situation before the results presented at this conference 
is summarized by the right plot in Fig.~\ref{fig:DeltaDs}, where the 
SM is about $3\sigma$ off. After the inclusion of the high-statistics data 
on $S_{\psi\phi}$ by both CDF and D0 announced at this conference~\cite{Borissov},
the deviation is expected to be slightly less than $2\sigma$
(an official combination is not yet available).

In this case the error is purely of experimental nature, and substantial 
improvement is expected in the near future, not only by the Tevatron 
but also by LHCb. Similarly to the case of  $\Delta_d$, also
the best fit value of $\Delta_s$ indicates a non-vanishing NP phase.
However, in the $B_s$ case the relative deviation from
the SM is substantially larger. If confirmed, this non-universal
pattern of deviations from the SM in $B_s$ and $B_d$ mixing
would provide a very powerful 
tool to discriminate possible NP models (see Sect.~\ref{sect:NP}).

\subsection{The $B\to \tau\nu$ decay}

The purely leptonic $B\to \ell\nu$ decays are particularly interesting 
for two main reasons. On the one hand they are theoretically very clean:
all hadronic uncertainties are confined to the $B$ meson decay 
constant ($f_B$), which can be computed reliably using Lattice QCD.
On the other hand, the strong helicity suppression makes them particularly 
sensitive probes of physics beyond the SM, especially of a 
non-standard Higgs sector. 

The $\tau$  channel is the only decay mode of this type observed so far. 
The expectation for the branching ratio within the SM has the 
following simple expression,
\be
\cB(B\to \tau \nu)^{\rm SM} = 
\frac{G_{F}^{2}m_{B}m_{\tau}^{2}}{8\pi}\left(1-\frac{m_{\tau}^{2}}
{m_{B}^{2}}\right)^{2}f_{B}^{2}|V_{ub}|^{2}\tau_{B}~.
\ee
Using the best fit value of $|V_{ub}|$ from global CKM fits,
and combining both direct lattice QCD constraints and 
indirect constrains from global CKM fits
on $f_B$~\cite{Bona:2009cj}, the UTfit collaboration obtains~\cite{Bevan:2010gi}
$\cB(B\to \tau \nu)^{\rm SM} = (0.79 \pm 0.07)\times 10^{-4}$.
This is substantially lower with respect to the current 
experimental world average,
$\cB(B\to \tau \nu)^{\rm exp} = (1.68 \pm 0.31)\times 10^{-4}$,
with a statistical significance of a deviation 
from the SM close to $3\sigma$. Even taking into account 
the more conservative estimate of the SM error quoted in~\cite{Bhattacherjee:2010ju},
the deviation exceed the $2\sigma$ level.

Beside possible experimental improvements on $B\to \tau \nu$, 
an important
ingredient to improve the significance of this SM test is the determination of $|V_{ub}|$ (that is relevant also for the $\sin2\beta$ problem 
discussed before).
Both these goals are possible at super-$B$ factories. As far as $|V_{ub}|$ is concerned, 
this could be systematically improved in the future with more precise experimental 
data on $B\to\pi \ell\nu$ combined with Lattice QCD results on the  $B\to \pi$ form-factor. 
Similarly to the approach presently adopted for the determination of $|V_{us}|$, the 
most promising strategy is the experimental determination of the 
kinematical dependence of the $B\to \pi$ form-factor  
combined with Lattice data to fix its overall normalization
(see e.g.~\cite{Antonelli:2009ws}).

\section{Possible beyond-the-SM explanations of these ``anomalies''}
\label{sect:NP}

None of the deviations from the SM discussed above 
has a high statistical significance; however, it is tantalizing 
to interpret them as possible hints of physics beyond the SM. 
Several attempts in this direction have been made in the recent 
literature. In the following I will focus on three classes 
of models where there has been
considerable activity in the last few months, and which are quite interesting
because of clear correlations among various observables:
i) the two Higgs doublet model (2HDM) with MFV and 
flavour-blind CPV phases; ii) an effective theory with right-handed (RH) currents; 
ii) the SM model with four generations of quarks.
None of these set-up represents a complete ultraviolet (UV) completion of the SM;
however, all of them can be viewed as ``simple'' effective theories which
could arise as the low-energy limit of more ambitious and more complete
theories.

\subsection{The ${\rm 2HDM}_{\overline{\rm MFV}}$ framework}

Two or more Higgs doublets are naturally 
expected in several UV completions of the SM, such as its minimal supersymmetric 
extension (MSSM). On general grounds, multi-Higgs models
suffer of too-large FCNCs, unless some extra protection mechanism is 
invoked (see e.g.~\cite{Botella:2009pq,Pich:2009sp,Gupta:2009wn,Buras:2010mh}
for a recent discussion). As already mentioned, an 
elegant way to justify the smallness of deviations from the SM in  
flavour-changing observables is provided by the MFV  hypothesis.
The ${\rm 2HDM}_{\overline{\rm MFV}}$ framework, as defined in~\cite{Buras:2010mh},
is nothing but the most general 2HDM (with two Higgses of hypercharge $\pm1/2$)
compatible with the MFV principle (flavour symmetry broken only by two Yukawa couplings),
with possible new flavour-blind CPV phases. 

\begin{figure}[t]
\begin{center}
\hskip -0.1 true cm
\raisebox{0.7 true cm}{\includegraphics[width=68mm]{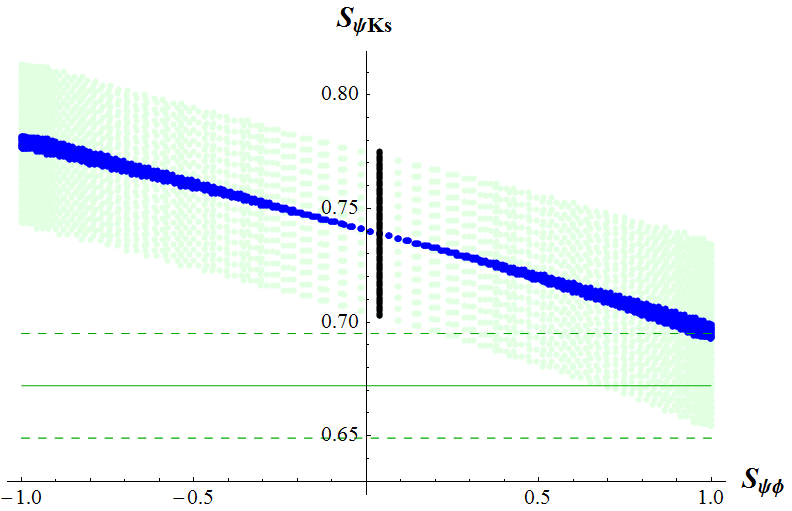}}
\hskip 0.3 true cm
\includegraphics[width=60mm]{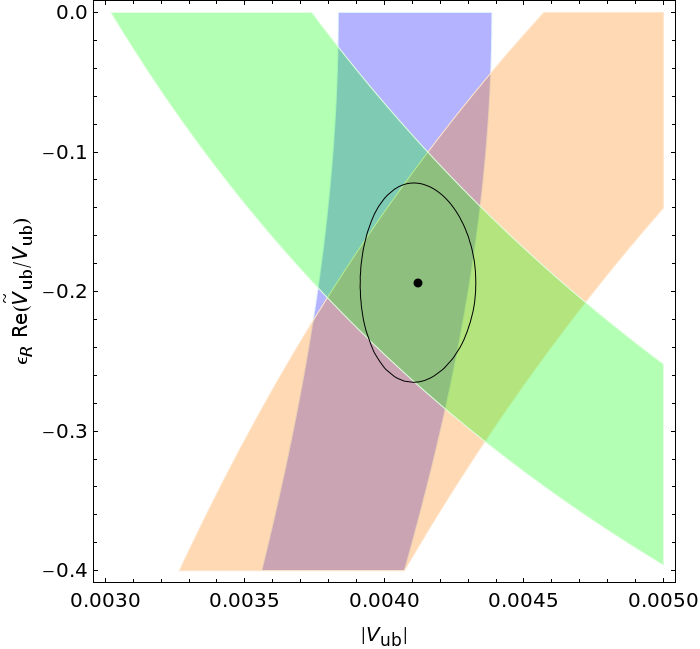}
\vskip  -5.6 cm 
\hskip   8.2 cm 
{\footnotesize ${}_{B\to \pi \ell \nu} \qquad {}_{B\to X_u \ell \nu} \qquad {}_{B\to \tau\nu}$ }
\vskip   5.2 cm 
\caption{\label{fig:sin2b} 
Left: correlation between $S_{\psi K_S}$ and $S_{\psi\phi}$ in the ${\rm 2HDM}_{\overline{\rm MFV}}$~\cite{Buras:2010mh} (the horizontal lines denote the experimental data, 
the black vertical line is the SM prediction). Right: constraints on $|V_{ub}|$ and 
the relative contribution from RH currents
from $B\to \pi \ell \nu$ (green), $B\to X_u \ell \nu$ (blue), and
$B\to \tau\nu$ (orange)~\cite{Buras:2010pz}. }
\end{center}
\end{figure}

This set-up is quite effective in suppressing FCNCs 
to a level consistent with experiments in most cases, 
leaving open the possibility of sizable non-standard effects 
only in specific observables sensitive to  Higgs-mediated 
FCNCs. The effective coupling controlling 
down-type Higgs-mediated FCNCs 
($d_R^i \to d_L^j H$) is suppressed both by the CKM combination 
$V_{ti}^* V_{tj}$ (similarly to FCNC amplitudes in the SM) and by the 
Yukawa coupling of the right-handed quark involved. 
This double suppression 
mechanism implies a well-defined pattern of deviations from the SM 
in $\Delta F=2$ amplitudes such that the largest corrections are
expected in $B_s$ mixing~\cite{Kagan:2009bn}.
Moreover, once the free parameters of the model are tuned to accommodate a large and 
positive $S_{\psi\phi}$, a small negative correction to $S_{\psi K}$
is automatically implied (see Fig.~\ref{fig:sin2b} left), 
with the ratio of the CPV phases in $\Delta_s$ and $\Delta_d$ 
unambiguously linked to $m_d/m_s$~\cite{Buras:2010mh}.  
This predictive pattern of deviations from the SM is 
quite interesting in view of the present data on 
$B_{s,d}$ mixing illustrated in the previous Section 
(see Fig.~\ref{fig:DeltaDs}).

Within an effective theory approach, the region of parameters
necessary to fit a large CPV phase in $B_{s}$ mixing 
is quite natural (heavy Higgs masses of the order of a few 100~GeV).
The new CPV phases are also compatible with present bounds from 
electric dipole moments (edms)~\cite{Buras:2010zm}.
However, in order to generate a large CPV phase in $B_{s}$ mixing 
a specific pattern of interference of effective operators 
with several Yukawa insertions is needed~\cite{Buras:2010mh,Ligeti:2010ia}.
This is not possible in the standard MSSM, but it could be realized 
in different underlying models, such as the up-lifted MSSM~\cite{Dobrescu:2010rh}.
The ``smoking-gun'' of this framework are $\cB(B_{s,d}\to \ell^+\ell^-)$
largely enhanced over their SM expectations, with 
$\cB(B_{s}\to \ell^+\ell^-)/\cB(B_{d}\to \ell^+\ell^-)$ fixed to its
SM value, and edms close to their present bounds~\cite{Buras:2010zm}.

\subsection{Right-handed currents}

One of the main properties of SM is the left-handed nature 
of the flavour-violating charged-current interactions.
This property could well be modified in extensions of the SM, 
such as left-right symmetric models, or even Higgs-less models.
A recent phenomenological interest in flavour-violating RH charged currents
originates from the tensions between inclusive and exclusive
determinations of $|V_{ub}|$. As pointed out in particular 
in~\cite{Crivellin:2009sd},  the presence of RH  
currents could either remove or significantly weaken these tensions.

Beside a clear benefit in $b\to u\ell \nu$  transitions
(see the right panel in Fig.~\ref{fig:sin2b}), it is interesting 
to understand if RH currents are compatible with constraints from 
other processes and where else they could show up. This problem can been 
addressed in general terms by means of an effective theory approach.
In particular, in Ref.~\cite{Buras:2010pz} it has been considered an effective 
theory based on the underlying $SU(2)_L \times SU(2)_R \times U(1)_{B-L}$ 
global electroweak symmetry, with the same low-energy particle content
as in the SM, and  with a left-right symmetric flavour group
broken only by two Yukawa couplings 
(RHMFV framework). A central role in this theory is played by a new unitary 
matrix ($\tilde V$) that controls flavour-mixing in the RH sector. 
This matrix, and the overall strength of RH currents ($\varepsilon_R$), 
can be constrained by all available data on semileptonic decays. Interestingly, a
sizable contribution to $b\to u$ (Fig.~\ref{fig:sin2b} right)
is not in contradiction with
the tight constraints on  $s\to u$ and  $d\to u$, 
and the effective scale of RH interactions turns out to be 
around 3 TeV.

As far as FCNCs are concerned, the most interesting 
implications of the RHMFV framework  
can be listed as follows: it is possible to generate a large 
CPV phase in $B_s$-mixing (as suggested by data); however, if this 
condition is fulfilled, no sizable effects in $B_d$-mixing
are expected. Moreover,  RH currents imply a  $\cO(10^{-3})$ deviation 
in the determination of $|V_{us}|$ from $K\to \pi\ell\nu$
and $K\to \ell\nu$ decays, that is close to the present experimental
sensitivity. Finally, non-standard contributions to
$B \to \{X_s,K, K^*\} \nu\bar \nu$ and
$K\to\pi\nu\bar\nu$ decays can be significant and, 
if present, the deviations from the SM in these decays 
would exhibit a well-defined pattern of correlations~\cite{Buras:2010pz}.

\subsection{Fourth generation}
The addition of a fourth generation of quarks and leptons to the SM 
is one of the simplest extensions of the SM one can conceive (SM4 framework).
The SM4 is not particularly interesting by itself; however, on the one
hand it is allowed by electroweak data, and actually it improves the 
quality of the electroweak fit~\cite{Holdom:2009rf}, on the other hand it provides 
a simple tool to analyse the impact of the mixing between SM and heavy fermions 
which is expected in several more complete extensions of the SM. 

In the last few years a renewed interest in the 
SM4 has been triggered by the possible impact in flavour-physics observables 
(see Ref.~\cite{Hou:2005hd,Bobrowski:2009ng,Soni:2010xh,Buras:2010pi}
and references therein). In the quark sector the model contains seven 
new parameters: two heavy quark masses, three new mixing angles and 
two new CPV phases. The quark masses are bounded to be below $\sim 600$~GeV
by the perturbativity of the corresponding Yukawa couplings, and the  
mass splitting is tightly constrained by electroweak data.
Global fits of the new flavour mixing parameters have been
performed in~\cite{Soni:2010xh,Buras:2010pi}. The main results 
can be summarized as follows: i) the tensions of the SM in $B_{s}$
and  $B_{d}$ mixing can both be solved; ii) a large CPV phase in 
$B_{s}$ mixing necessarily implies a suppression of the time-dependent 
CPV asymmetries in $b\to s$ penguin-type modes, $B\to \phi K$ and 
$B\to \eta^\prime K$, in agreement with data (as noted first in \cite{Hou:2005hd}), 
it also implies a sizable enhancement of $\cB(B_s \to \mu^+\mu^-)$, 
with  $\cB(B_d \to \mu^+\mu^-)$ close to its SM value, a clear 
non-standard pattern that can be tested in the near future;
iii) a large CPV phase in $B_{s}$ mixing also implies a tension between
the prediction of $\epsilon^\prime/\epsilon$ and its measurement, 
although this effect is hidden by the current uncertainty 
on $K\to \pi\pi$ hadronic matrix elements; 
iv) the rare $K\to\pi\nu\bar\nu$ decays can be largely enhanced over their SM predictions.

\section{Experimental challenges for the near future}

Current ``anomalies'' are certainly interesting, but we cannot exclude 
they will all disappear with higher statistics. Indeed they are not the most 
natural expectations in the most ``conservative'' beyond-SM scenarios, 
such as models with MFV and no extra CPV phases. On the other hand,
there are a few other channels where we can expect sizable deviations 
from the SM even in  ``conservative'' beyond-SM scenarios, and for 
which we can expect precise experimental results in the near future.
Among the most promising channels for the near future it is worth to 
mention the following three cases: i) the search for lepton flavour violation (LFV) in charged leptons; ii) the very rare 
FCNC decays $K\to \pi \nu\bar \nu$; iii) the helicity- suppressed 
FCNC $B$ decays $B \to \ell^+\ell^-$.

\begin{figure}[t]
\begin{center}
\hskip -0.3 true cm
\includegraphics[width=60mm]{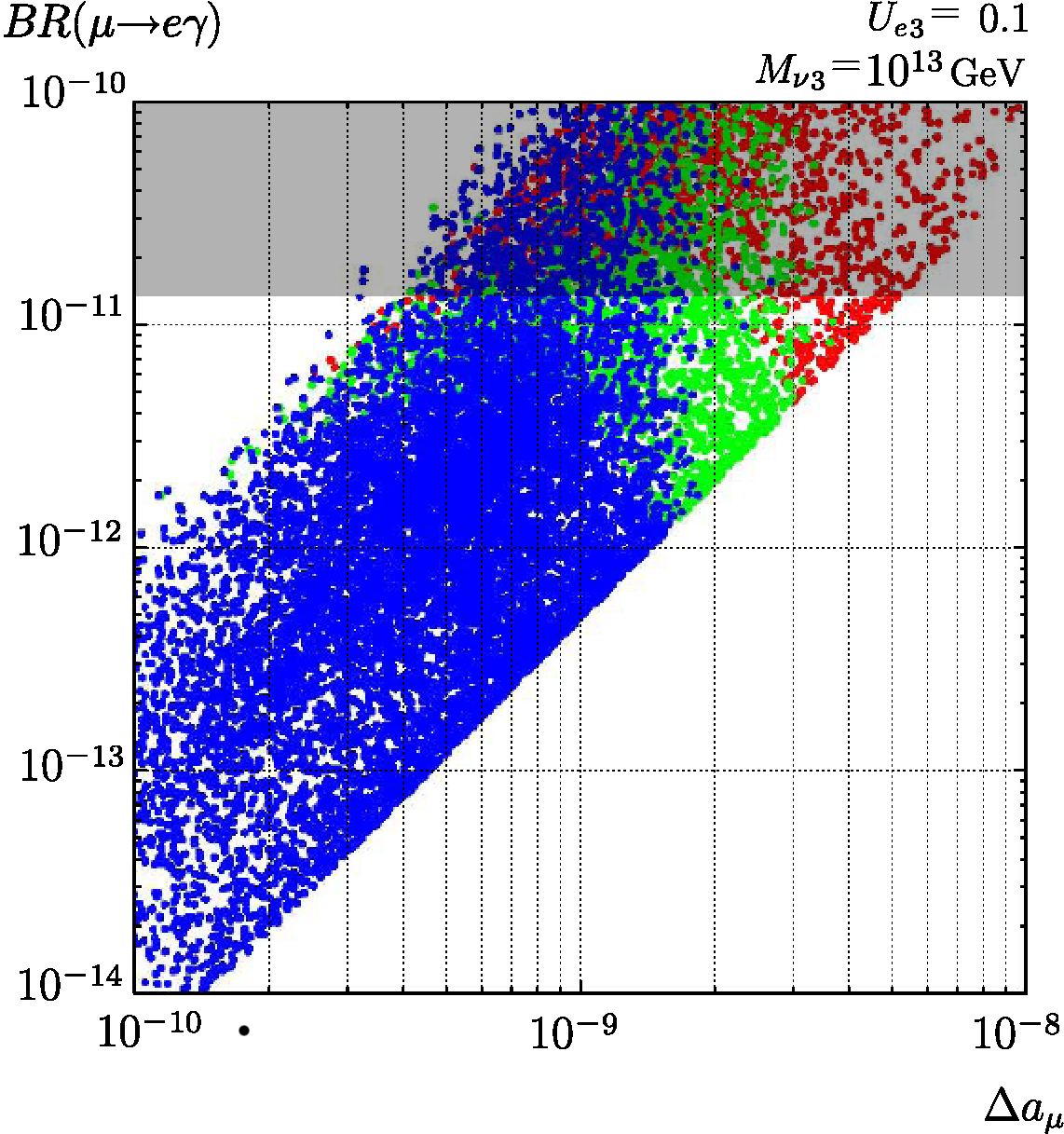}
\hskip 0.5 true cm
\raisebox{1.0 true cm}{\includegraphics[width=68mm]{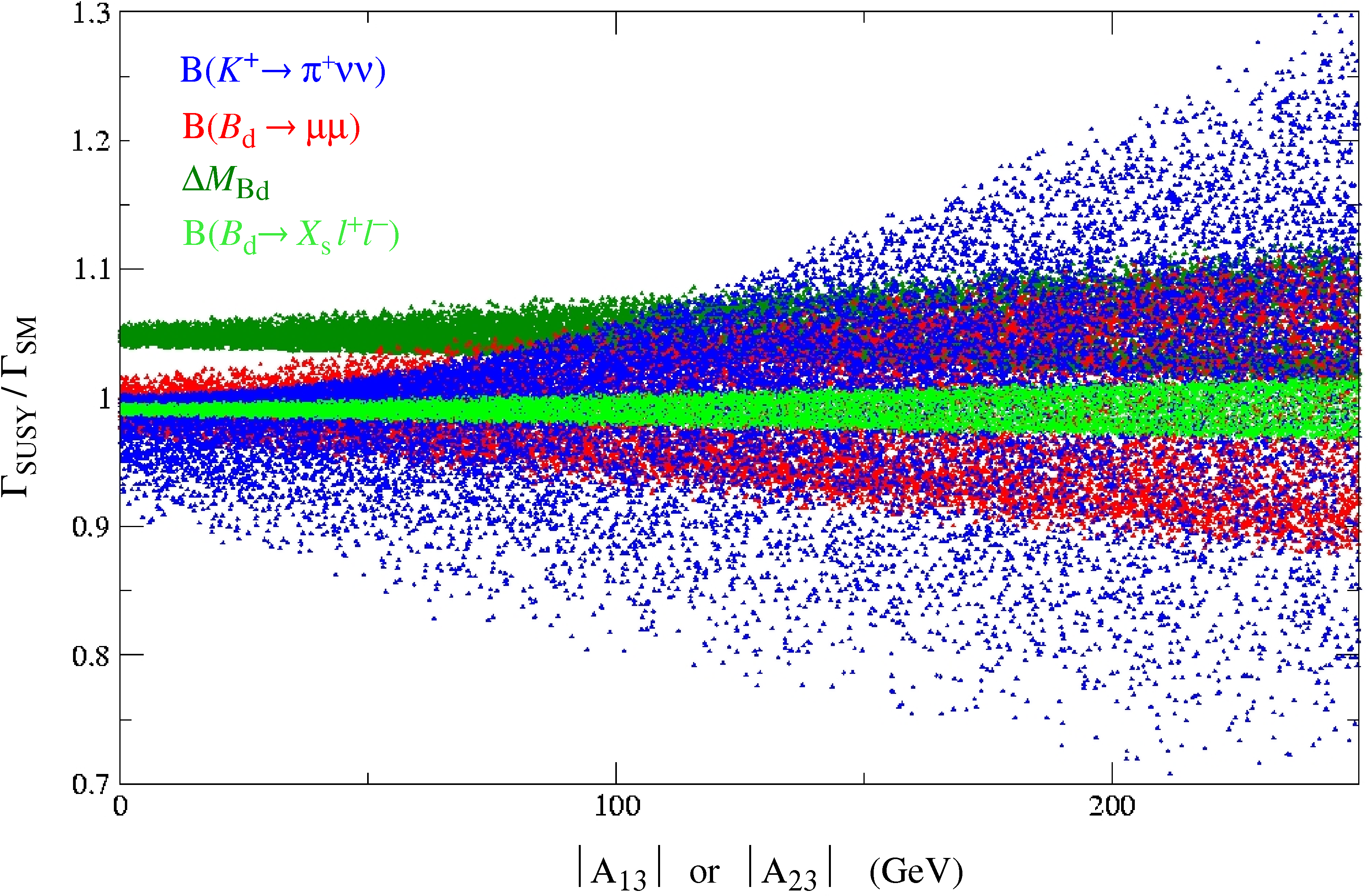}}
\caption{\label{fig:rareK} 
Left: correlation between $\cB(\mu\to e\gamma)$ and $\Delta a_\mu$ in the MSSM with heavy 
RH neutrinos~\cite{Hisano:2009ae}.
Right: dependence of various FCNC 
observables (normalized to their SM value) on the up-type trilinear terms $A_{13}$ and $A_{23}$
(generically denoted by $A_{13}$) in the MSSM with non-minimal $A$ terms~\cite{Isidori:2006qy}. }
\end{center}
\end{figure}

\smallskip

The search for LFV in charged leptons is probably the most interesting
goal of flavour physics in the next few years. The observation of 
neutrino oscillations has clearly demonstrated that lepton flavour is 
not conserved; however, the smallness of neutrino masses provides a 
strong indication that neutrinos are generated by an underlying dynamics 
that violates also the total lepton number.
The question is if LFV effects can be visible also in other 
sectors of the theory, or if we can observe  LFV in processes 
which conserve the total lepton number. The most promising 
low-energy channel is $\mu\to e\gamma$,
currently under investigation at MEG~\cite{MEG}. 
On general grounds, if the breaking of the 
total lepton number occurs at a very high energy scale ($\Lambda_{\rm LN} > 10^{12}$~GeV), 
as expected by the smallness of neutrino masses,
and the theory has new degrees of freedom carrying lepton-flavour quantum numbers around 
the TeV scale ($\Lambda_{\rm LFV} < 10^4$~GeV), then
$\mu\to e\gamma$ should be visible. Indeed, employing an 
effective theory approach with a minimal breaking of lepton flavour,
we find~\cite{Cirigliano:2005ck}
\be
\cB(\mu\to e\gamma) \approx 10^{-13} 
\left( \frac{\Lambda_{\rm LN}}{10^{13}~{\rm GeV}}\right)^4 
\left( \frac{10^4~{\rm GeV}}{\Lambda_{\rm LFV}}\right)^4~.
\ee
A typical concrete example where this occurs is the 
MSSM with heavy right-handed neutrinos, where 
renormalization-group effects generate
LFV entries in the left-handed slepton mass matrices at the TeV scale~\cite{Borzumati:1986qx}.
Once non-vanishing LFV entries in the slepton mass matrices 
are generated, LFV rare decays are naturally induced by
one-loop diagrams with the exchange of gauginos and sleptons.
The flavour-conserving component of the 
same diagrams induces a non-vanishing contribution to the 
anomalous magnetic moment of the muon, $\Delta a_\mu=(g_\mu-g^{\rm SM}_\mu)/2$.
As shown in Fig.~\ref{fig:rareK} (left), a strong link between 
these two observable naturally emerges (see e.g.~\cite{Hisano:2009ae}).
In this context, the value $\Delta a_\mu =\cO(10^{-9})$, 
presently indicated by detailed analyses of $g_\mu$~\cite{Davier:2010nc}, 
reinforce the expectation of 
$\mu\rightarrow e\gamma$  within the reach of the MEG experiment.

\smallskip

Among the many rare $K$, $D$, and $B$  decays, 
the $K\to \pi \nu\bar\nu$ modes are unique since their SM branching ratios
can be computed to an exceptionally high degree of precision, not
matched by any other FCNC processes involving quarks
(see~\cite{Brod:2010hi} for the most updated SM predictions).
It is then not surprising that these processes 
continue to raise a strong theoretical interest, both within 
and beyond the SM (see e.g.~\cite{Buras:2010pz,Blanke:2009am,Bauer:2009cf}). 
Because of the strong suppression of 
the $s \to d$ short-distance amplitude in the SM [$V_{td}V_{ts}^* =\cO(10^{-4})$],
rare $K$ decays are the most sensitive probes of possible deviations from the  
strict MFV ansatz. An illustration of this statement is given 
in Fig.~\ref{fig:rareK} (right), where the expectations of 
$\cB(K^+\to\pi^+\nu\bar\nu)$ in the MSSM with 
non-minimal $A$ terms is shown in comparison with possible deviations from
the SM in other rare-decay observables. As can be seen, $\cB(K^+\to\pi^+\nu\bar\nu)$ 
is the most sensitive probe of this class of models.

\begin{figure}[t]
\begin{center}
\includegraphics[width=65mm]{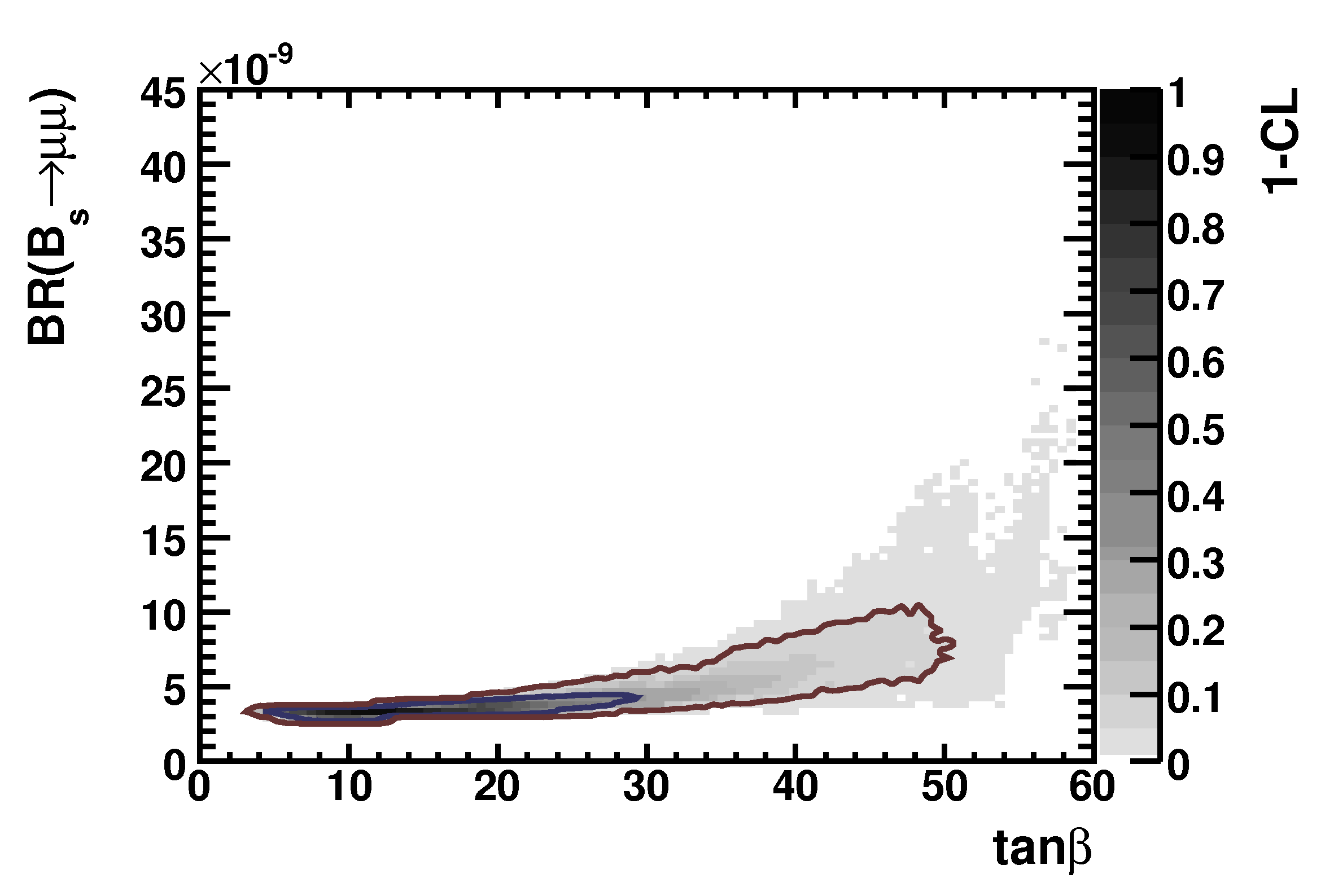}
\includegraphics[width=65mm]{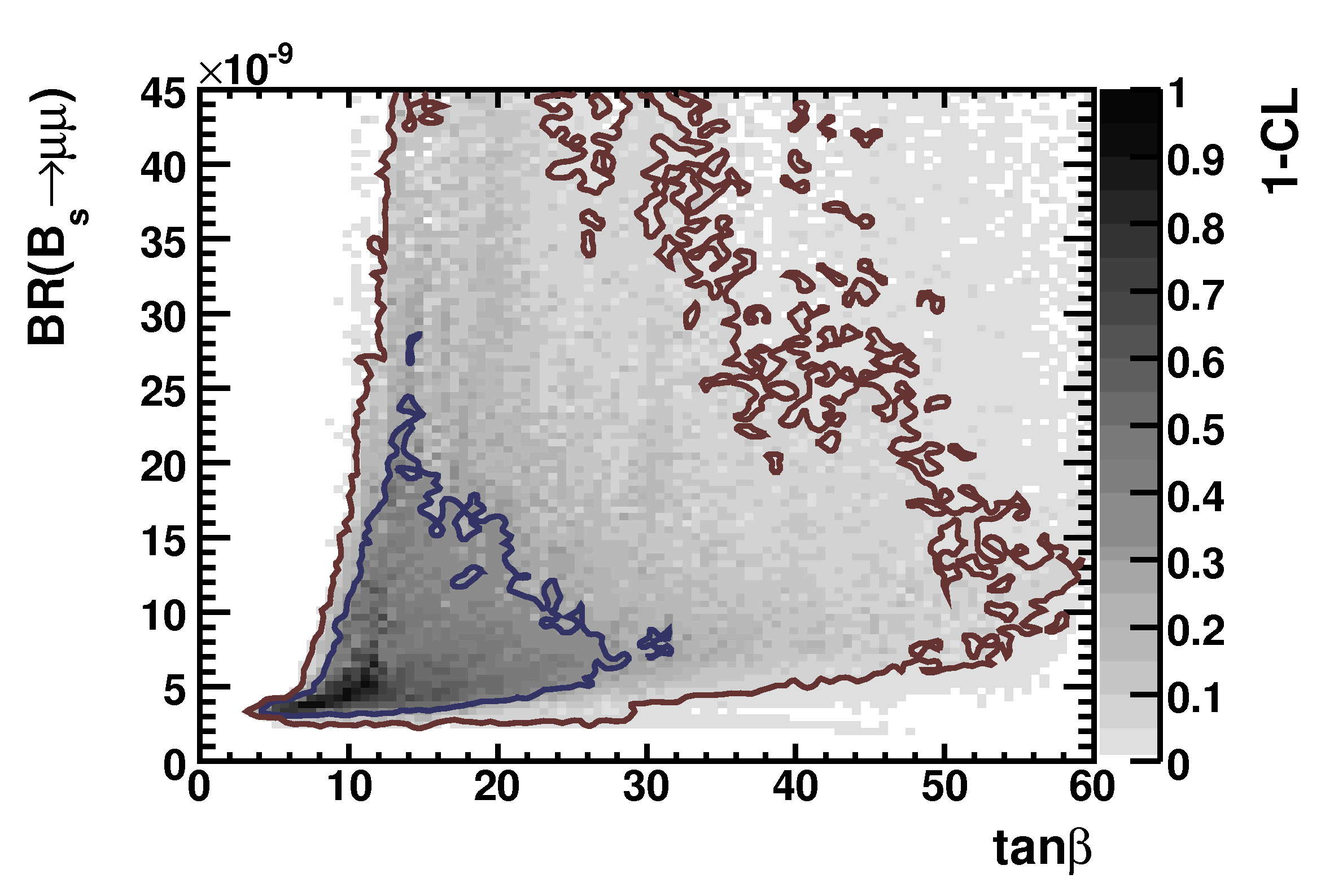}
\caption{\label{fig:future}
The correlation between $\cB(B_s\to\mu^+\mu^-)$ and $\tan\beta$
in the constrained MSSM (left panel) and in the MSSM with non-universal Higgs
soft-mass terms (right panel)~\cite{Buchmueller:2009fn}.
In both panels the CL is obtained combining all the available (indirect) 
constraints on the model~\cite{Buchmueller:2009fn}.}
\end{center}
\end{figure}

\smallskip

The rare decays $B \to \ell^+\ell^-$ are particularly interesting because 
in the SM suffer both a strong GIM suppression (being FCNC transitions) 
and a strong helicity suppression (being two-body leptonic decays). 
This doubly-suppressed structure make them ideal probes of possible 
Higgs-mediated FCNC amplitudes in models with more than one Higgs 
doublet, such as the MSSM. Even in the most restricted versions 
of the MSSM (with MFV and with no new CPV phases),  $B \to \ell^+\ell^-$
decays could be largely enhanced over the SM expectations 
if $\tan\beta$  (the ratio of the two Higgs vevs) is large.
The leading non-SM amplitude contributing to $B\to \ell^+\ell^-$ is
generated by the heavy neutral Higgs exchange ($B\to A,H \to \ell^+\ell^-$). 
However, since the effective FCNC coupling of the neutral 
Higgs bosons is generated at the quantum level, the amplitude 
has a strong dependence on other MSSM parameters in addition to  
$M_{A,H}$ and $\tan\beta$. In particular, a key role is played by $\mu$ and 
the up-type trilinear soft-breaking term ($A_U$),
which control the strength of this effective vertex. 
The leading parametric dependence of the scalar FCNC amplitude 
from these parameters is 
\bea
\cA_{\rm Higgs}( B \to \ell^+\ell^-) \propto
 \frac{m_b m_\ell}{M_A^2}
 \frac{\mu A_U}{M^2_{\tilde q}} \tan^3\beta 
\times f_{\rm loop}\!\!\!\!  \no
\eea
For $\tan\beta \gsim 30$ and $M_A \lsim 0.5$~TeV
the neutral-Higgs contribution can easily lead to $\cO(10)$ enhancements 
of $\cB(B_{s,d} \to \ell^+ \ell^-)$  over the SM expectations.
Most important, these decays represent a very 
useful tool to determine a combination of MSSM parameters that 
would help in discriminating different versions of the model
(see e.g.~Fig.~\ref{fig:future}). At present the most significant 
constraints are obtained from  $\cB(B_{s} \to \mu^+ \mu^-)$, 
where the experimental upper limit,
$\cB(B_{s} \to \mu^+ \mu^-) < 5.8\times 10^{-8}$~\cite{:2007kv},
is less than 20 times the SM expectation, 
$\cB(B_{s} \to \mu^+ \mu^-)_{\rm SM} = (3.2 \pm 0.2) \times 10^{-9}$.

\section{Conclusions}

The origin of flavour remains, to a large extent, an open problem.
However, a significant progress has been achieved in the 
phenomenological investigation of the  sources of flavour 
symmetry breaking accessible at low energies. This investigation 
has allowed to set very stringent constraints on various extensions
of the SM, ruling out models with significant  misalignments from the 
SM Yukawa couplings at the TeV scale.

What we learned so far does not imply we cannot 
see some deviation from the SM in low-energy process 
in the near future. 
A few interesting anomalies in the CKM picture have started to emerge.
Some of them will go away with more data, but others 
may well be the first signals of  new physics at the TeV scale.
Even more interesting are the prospects of finding deviations from 
the SM, or to set very powerful constraints on physics beyond the SM,
in clean processes such as $\mu\to e\gamma$, 
$B\to \ell^+\ell^-$, and $K\to\pi\nu\bar\nu$, where a significant
experimental progress is expected in the next few years. 
As briefly outlined in this talk (see Ref.~\cite{Buras}
for a more extensive discussion), once some clear deviation 
from the SM will be established, the key tool to make progress in this field 
is to identify correlations among different non-standard effects that 
can reveal the flavour-breaking pattern of the new degrees of freedom.

\section*{Acknowledgment}
This talk is dedicated to the ``father'' of flavour physics, 
Nicola Cabibbo, who sadly passed away a few weeks 
after the end of this conference.

\end{document}